\documentclass[manuscript]{emulateapj}
\slugcomment{Accepted for publication in The Astrophysical Journal}

\shorttitle{Globular clusters in NGC 474}
\shortauthors{Lim et al.}

\begin{document}


\title{Globular Clusters As Tracers of Fine Structure \\
in the Dramatic Shell Galaxy NGC 474}


\author{Sungsoon Lim\altaffilmark{1,2}, Eric W. Peng\altaffilmark{1,2}
, Pierre-Alain Duc\altaffilmark{3}
, J{\'e}r{\'e}my Fensch\altaffilmark{3}
, Patrick R. Durrell\altaffilmark{4}
, William E. Harris\altaffilmark{5}
, Jean-Charles Cuillandre\altaffilmark{6}
, Stephen Gwyn\altaffilmark{7}
, Ariane Lan{\c c}on\altaffilmark{8}
, R{\'u}ben S{\'a}nchez-Janssen\altaffilmark{9}}

\altaffiltext{1}{Department of Astronomy, Peking University, Beijing, 100871, China; slim@pku.edu.cn, peng@pku.edu.cn}
\altaffiltext{2}{Kavli Institute for Astronomy and Astrophysics, Peking University, Beijing, 100871, China}
\altaffiltext{3}{Laboratoire AIM Paris-Saclay, CNRS/INSU, Universit{\'e} Paris Diderot, CEA/IRFU/SAp, F-91191 Gif-sur-Yvette Cedex, France}
\altaffiltext{4}{Department of Physics and Astronomy, Youngstown State University, Youngstown, OH 44555, USA}
\altaffiltext{5}{Department of Physics \& Astronomy, McMaster University, Hamilton, ON, Canada}
\altaffiltext{6}{Canada–France–Hawaii Telescope Corporation, Kamuela, HI 96743, USA}
\altaffiltext{7}{Herzberg Institute of Astrophysics, National Research Council of Canada, Victoria, BC, V9E 2E7, Canada}
\altaffiltext{8}{Observatoire astronomique de Strasbourg, Universit{\'e} de Strasbourg, CNRS, UMR 7550, 11 rue de l’Universit{\'e}, F-67000 Strasbourg, France}
\altaffiltext{9}{STFC UK Astronomy Technology Centre, The Royal Observatory Edinburgh, Blackford Hill, Edinburgh, EH9 3HJ, UK}







\begin{abstract}
Globular clusters (GCs) are some of the most visible tracers of the merging and accretion history of galaxy halos. Metal-poor GCs, in particular, are thought to arrive in massive galaxies largely through dry, minor merging events, but it is rare to see a direct connection between GCs and visible stellar streams. NGC 474 is a post-merger early-type galaxy with dramatic fine structures made of concentric shells and radial streams that have been more clearly revealed by deep imaging.
We present a study of GCs in NGC 474 to better establish the relationship between merger-induced fine structure and the GC system.
We find that many GCs are superimposed on visible streams and shells, and about $35\%$ of GCs outside $3R_{\rm e,galaxy}$ are located in regions of fine structure.
The spatial correlation between the GCs and fine structure is significant at the $99.9\%$ level, showing that this correlation is not coincidental. 
The colors of the GCs on the fine structures are mostly blue, and we also find an intermediate-color population that is dominant in the central region, and which will likely passively evolve to have colors consistent with a traditional metal-rich GC population.
The association of the blue GCs with fine structures is direct confirmation that many metal-poor GCs are accreted onto massive galaxy halos through merging events, and that progenitors of these mergers are sub-$L^\star$ galaxies. 

\end{abstract}





\section{Introduction}

In the current view of galaxy formation, galaxies are formed by the hierarchical assembly of smaller sub-components. In massive galaxies, this is visible in the outer regions, which are expected to grow through multiple dry mergers of low-mass, relatively metal-poor galaxies (e.g. \citealp{Naa09,Ose10}). This merging is also thought to be the origin of metal poor globular clusters in massive early-type galaxies \citep{Cot98,Li14}. 

Globular clusters (GCs) are powerful tools to study the assembly of their host galaxies. They are formed in intense star forming events, and can survive a Hubble time. They are also abundant in galaxies, 
across many orders of magnitude in mass. Conveniently, the relatively high surface brightness of GCs 
makes them observable in distant galaxies. 

Metal-rich GCs are formed in major, gas-rich mergers of relatively massive systems, where the star-forming gas is more enriched. Many theoretical studies support this formation mechanism of metal-rich GCs (e.g. \citealp{Bou08,Kru12,Ren13,Li14,Ren15}). 
Several observational studies also support this formation scenario (e.g. \citealp{Whitmore93,Sik06,Tra14}).

\begin{figure*}
\epsscale{1.0}
\plotone{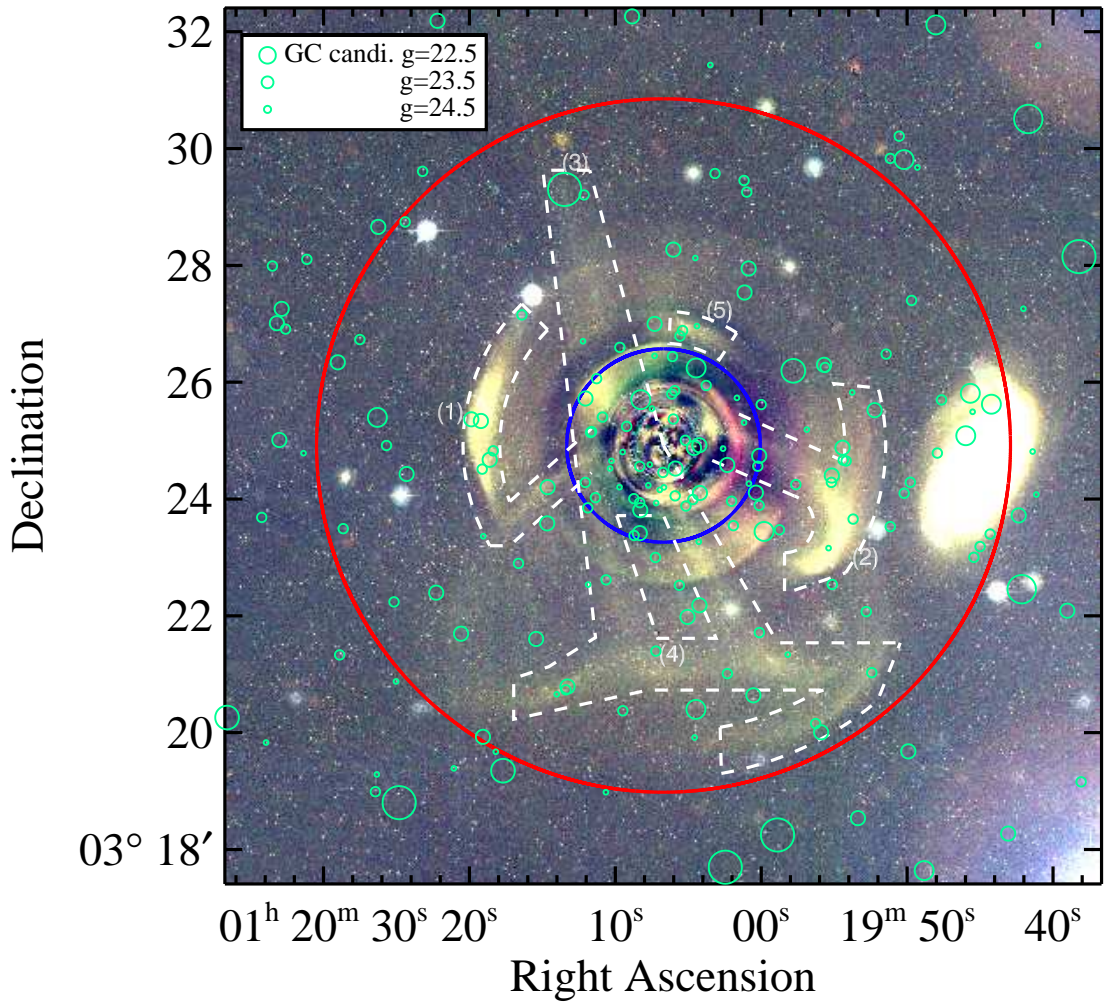}
\caption{Spatial distribution of GC candidates centered on NGC~474. The coordinates are J2000. Small open circles represent GC candidates of NGC~474. Sizes show $g$ band magnitudes of GC candidates with the size scale displayed at the top left. Two large circles centered on the galaxy (inner and outer) show $3$ effective radii for the stars, and $2.5$ effective radii of the GC system, respectively. Dashed lines indicate the rough boundaries of fine structures, with numbers labeling the interesting substructures. The color image was made with $gri$ model-subtracted images. Field of View is $15\arcmin\times15\arcmin$ corresponding $135{\rm kpc}\times 135{\rm kpc}$ in NGC~474 distance. \label{spa}}
\end{figure*}

Metal-poor GCs, however, are thought to form in low-mass galaxies. When seen in halos of massive galaxies, these GCs are thought to be the visible remnants of minor merging events \citep{Cot98,Tonini13}. There is much indirect evidence that the metal-poor GCs in dwarfs are related to those in the halos of massive galaxies. The specific frequency ($S_{\rm N}$) of metal-poor GCs is as high in massive galaxies as it is in some dwarfs. In the Virgo cluster, there is evidence that $S_{\rm N}$ in dwarfs is higher when they are closer to the central galaxy, M87, indicating that they may contribute to the progenitor population of the M87 metal-poor GC population \citep{Pen08}. Moreover, these satellite dwarfs have the high [$\alpha$/Fe] seen in galaxy halo stars \citep{Liu16}. The azimuthal distributions and ellipticities of metal-poor GCs are also generally uncorrelated with those of their host galaxies \citep{Wan13,Par13}, suggesting a random merger history. The kinematics of metal-poor GCs and stars in the outskirts of galaxies are also different from that of the stars in the inner regions \citep{Pota13,Li15,Fos16}. 
While these observations are all suggestive that metal-poor GCs in massive galaxies come from accreted galaxies, it has been rare for the process to be observed directly. One of the few exceptions has been in the M31 PAndAS survey, which revealed two GCs associated with visible stellar substructures \citep{Hux14}. Another example is NGC~4651, a late-type spiral galaxy (Sc) that has GCs associated with currently observed streams and shells \citep{Fos14}. 
Above examples are late-type galaxies, and there is only one example of the giant elliptical galaxy, NGC~4365 that has GCs related to the stellar stream \citep{Blo12}.

Establishing the link between GCs and mergers in the halos of massive early-type galaxies is difficult because GCs are a sparse population, and stellar substructures from merging have very low surface brightness. Studies of GC systems in shell galaxies with HST \citep{Sik06} have not had the field-of-view necessary to probe the outermost substructures where we expect the GC population to be dominated by the accreted component.
The Mass Assembly of Early-Type Galaxies with their Fine Structures (MATLAS; \citealt{Duc15}) survey is the first systematic, wide-field imaging survey that goes deep enough in surface brightness to reveal merging substructure and simultaneously has a good image quality for unresolved sources to allow a detailed study of the corresponding GC systems. 
To demonstrate the potential of this data, in this paper, we present a study of GCs in the spectacular post-merger galaxy, NGC 474.

\section{Observations}
These observations were part MATLAS, a Canada-France-Hawaii Telescope (CFHT) Large Program to obtain deep, multi-band optical imaging for 260 ATLAS$^{\rm 3D}$ galaxies \citep{Cap11} using MegaCam on CFHT. 
The detailed strategies for dithering and stacking images are described in \citet{Duc15}.
The stacking, however, for the GC analysis in this paper was done slightly differently. While the background subtraction was done as described in that paper, MegaPipe \citep{Gwy08} was used to do a
high-precision astrometric and photometric calibration. The MegaPipe output image stacks are produced at the full CFHT resolution, whereas the earlier image stacks were binned on 3x3 pixel grid.

NGC~474 is one of most interesting galaxies in the MATLAS sample ($D=30.9$~Mpc; \citealt{Cap11}). 
It is well known as a shell galaxy \citep{Turn99,Sik07}, and the deep $u^*$, $g'$, $r'$, and $i'$-band imaging reveals very complex but clear fine structures made of concentric shells in addition to radial structures (Figure~\ref{spa}), particularly in the halo regions, where we can easily distinguish fine structures from the smooth stellar light. 
Accumulated exposure times are 4900s, 2415s, 2415s, and 3220s for $u^*$, $g'$, $r'$, and $i'$, respectively.

\begin{figure}
\epsscale{1.2}
\plotone{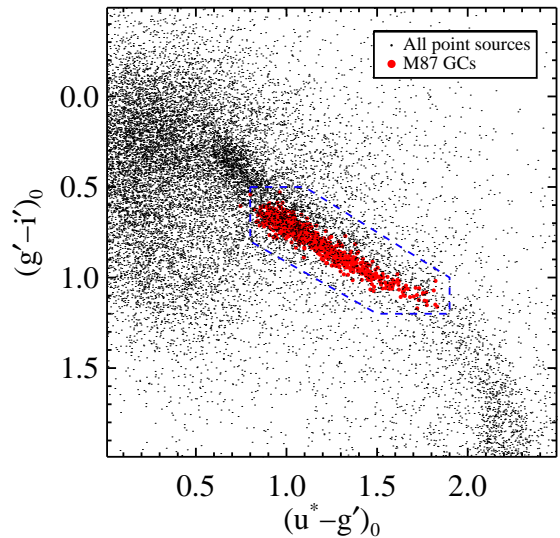}
\caption{$(u^*-g')_0$ vs. $(g'-i)_0$ color-color diagram of point sources. Small dots represent point sources in this study. Filled circles indicate colors of spectroscopically confirmed GCs of M87 in previous studies to guide location of GCs in the color-color diagram. \citep{Han01,Str11,Zha15}. We drew dashed polygon based on M87 GCs, and chose GC candidates in NGC 474 using this polygon \label{ccd}}
\end{figure}

We used Source Extractor \citep{Ber96} for source detection and photometry. Dual-image mode photometry was adopted, with the $r$-band image used as the detection image. 
We used circular apertures with various radii for estimating fluxes of sources. 
All aperture magnitudes are corrected to the 16 pixel diameter aperture magnitudes.  
These instrumental magnitudes are transformed to standard AB magnitudes by comparison with SDSS PSF magnitudes. 
The methods for source detection and aperture photometry are similar to those adopted in \citet{Dur14} and \citet{Liu15}. 
The foreground reddening toward NGC~474 is $E(B-V)=0.04$ \citep{Sch98}, and the corresponding extinction correction terms are included in the following analysis. 
At the distance of NGC 474, most of GCs are point-like sources, so we selected point sources using a concentration index with a magnitude limit of $m_{g',0} \leq 25$. At the distance of NGC~474, this magnitude limit should include roughly half of the GCs.
The concentration index is calculated by differences between 4 pixel diameter and 8 pixel diameter aperture magnitudes. 
Among the point-like sources, we selected GC candidates based on their colors in the $(u^*-g')_0-(g'-i')_0$ color-color diagram (Figure~\ref{ccd}).

\section{Results}


\begin{figure}
\epsscale{1.2}
\plotone{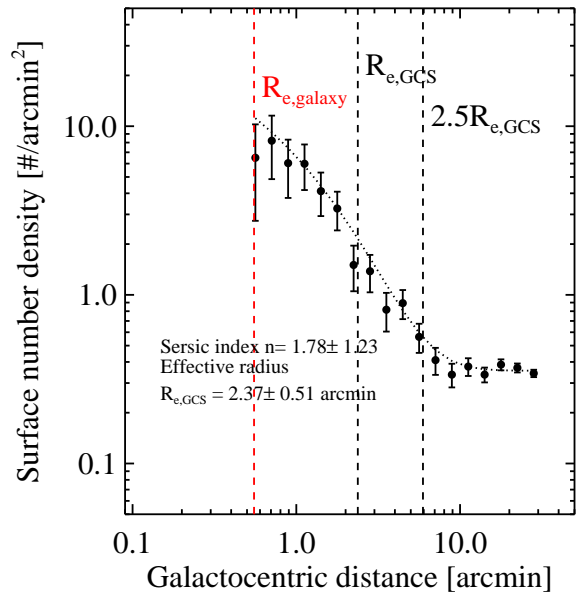}
\caption{Number density profile of GC candidatess. Filled circles with error bars show number density of GC candidates with binned area and its errors. The dotted line represent the best fit with a Sersi\'c funtion including background. Vertical dashed lines display effective radius of galaxy, effective radius of GC system, and 2.5 effective radius of GC system, respectively. \label{ndp}}
\end{figure}

\begin{figure}
\epsscale{1.2}
\plotone{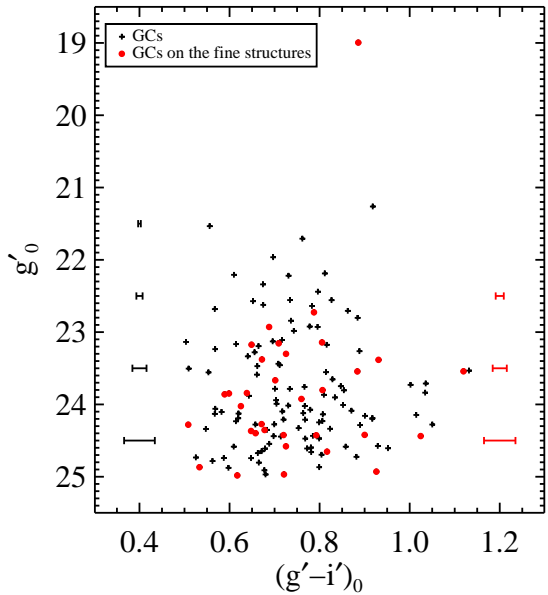}
\caption{$(g'-i)_0 - g_0$ color magnitude diagram of GC candidates within 2.5$R_{\rm e,GCS}$. Crosses and filled circles show GC candiddates off and on the fine structures, respectively. Error bars on the left side  and right side represent color errors of GC candidates off and on the fine structures, respectively. \label{cmd}}
\end{figure}

\begin{figure}
\epsscale{1.1}
\plotone{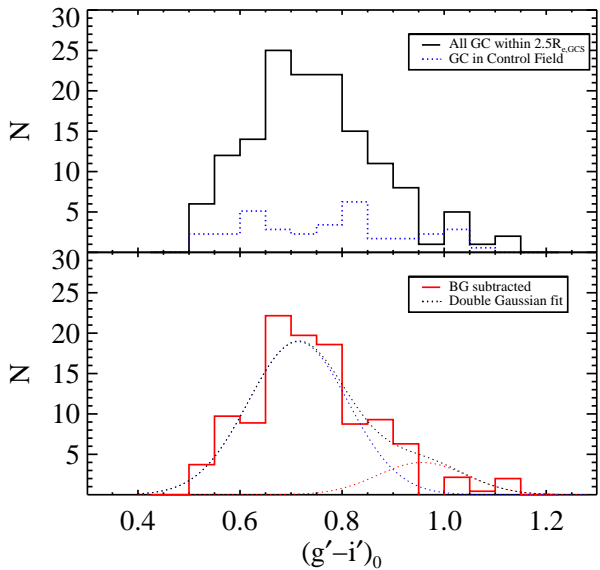}
\caption{Color histogram of GC candidates in NGC~474. $(Top)$ The solid line shows the color histogram for all GC candidates within $2.5$ effective radius of GC system. The dotted line (blue) displays the color histogram of background contamination (histogram of GC candidates in the control field, corrected for area). $(Bottom)$ Solid line (red) represents the background subtracted color histogram of GC candidates within $2.5$ effective radius of GC system. Dotted lines show the fitted Gaussian functions. \label{colh}}
\end{figure}

\subsection{Global Properties}

Figure~\ref{spa} shows the spatial distribution of GC candidates and fine structures in NGC~474. The inner region of NGC~474 suffers from incompleteness due to the bright stellar light and the difficulty of detecting GCs on the inner fine structures. Therefore, we only analyze the GC system beyond $1R_{\rm e,galaxy}$ ($0.55\arcmin$), out to $R=15\arcmin$. 
Using our GC selection in regions well away from the galaxy (beyond regions shown in Figure~\ref{spa}), we estimate the level of background contamination to be $0.03$~arcmin$^{-2}$.
We fit the azimuthally-averaged radial surface density profile of the GCs with a Sersi\'c profile using maximum likelihood estimation, and determined the uncertainties in the fitted parameters using the bootstrap. The 
effective radius and Sersi\'c index of the GC system are $R_{\rm e,GCS}=2\farcm37\pm0\farcm51$ ($21.1\pm4.5$~kpc at the distance of NGC~474) and $n=1.78\pm1.23$, respectively (Figure~\ref{ndp}).
The Sersi\'c index has a large uncertainty because it is sensitive to the lack of information in the central region. 
The effective radius of GC system is about 4 times larger than that of the stars. This relation is consistent with the result from \citet{Kar14}. For the analysis below, we chose GC candidates within $2.5R_{\rm e,GCS}$, represented by the largest circle (red) in Figure~\ref{spa}. To estimate contamination, we chose a control field between $5R_{\rm e,GCS}$ and $6R_{\rm e,GCS}$. There are 144 GC candidates within 2.5$R_{\rm e,GCS}$, and the estimated background contamination value for this area is about 34. The number density of sources that satisfied GC selection criteria in the control field gives the statistical estimate of the contamination. Table~\ref{tbl1} shows the list of GC candidates in 2.5$R_{\rm e,GCS}$.

Figure~\ref{cmd} shows the $(g'-i')_0 - g'_0$ color magnitude diagram of GC candidates within 2.5$R_{\rm e,GCS}$. GC candidates are divided by their location--off or on the fine structures. The brightest GC candidate is $g'_0\sim19$, but most GC candidates are fainter than $g'_0\sim 21$. Color ranges and errors are not location-dependent , but GC candidates on the fine structures tend to be fainter that those off the fine structures. The color errors are quite small ($< 0.1$) even with faint magnitude bin.

Figure~\ref{colh} shows the $(g'-i')_0$ GC color histograms. 
The color distribution of GC candidates ranges from $(g'-i')_0=0.5$~mag to $(g'-i')_0=1.15$~mag, and has a peak at $(g'-i')_0 \sim 0.75$~mag. 
We note that GC candidates redder than $(g'-i')_0=0.9$~mag are rare. 
Background contamination should not affect to shape of color distribution, as the color distribution of selected background objects is relatively flat. 
We fit the color histogram with Gaussian Functions using the Gaussian Mixture Modelling (GMM) code \citep{Mur10}. To subtract background contamination, we adopted the following steps. (1) We estimated the total number of contaminants, 34, based on the GC number density of the control field. (2) We randomly selected 34 GCs in the control field, and subtracted 34 GCs in the target region that have similar colors to the selected “GCs” from the control field. (4) The width for the two Gaussian functions are free parameters, and we ran GMM 1000 times with a random subtraction of background contamination. The GMM code provides the blue and red peaks of color distribution, and D value. The D value shows separation of two peaks relative to their width, and two-Gaussian fitting is meaningful when D $> 2$. The mean D value for 1000 GMM trials is 3.00 and its standard deviation is 0.36. It suggests that the color histogram of GCs in NGC 474 is better explained by a bimodal distribution than a unimodal distribution, although this is likely due to the tail of red GCs. Dust in NGC 474’s shells (e.g., \citealp{Sik07}) could also cause reddening of GC colors, but we see no correlation between the location of red GCs and position relative to shells.
The two Gaussian distributions have means at $(g'-i')_0=0.71\pm0.02 $ and $0.96\pm0.09$, respectively, and and these values are consistent with expected color peaks from the empirical relations between GC peak colors and galaxy luminosity \citep{Pen06}.

\subsection{Fine structures}
The fine structures of NGC 474 are very complex and have been discussed previously \citep{Turn99,Duc15}. 
Two relatively bright substructures are located at the east and west sides (1 and 2 in Figure ~\ref{spa}, respectively). 
Substructure (2) has a tail which seems to penetrate the central region of NGC 474. 
A tidal stream (3) crosses the main body of NGC 474 from the North to the South. 
Tidal streams are also found at the southern area (4). 
These streams (4) may be linked with the tail of the substructure (2). 
There is also a bright shell structure to the north of the central region (5). 
Many shell structures are seen in the inner regions of NGC 474, but some of them may be due to imperfect model subtraction. We ignore these structures for the purposes of this study.

To estimate the number and properties of GCs in the fine structures, we define an area that includes all regions described above, except for the central region ($R<1\farcm7 \approx 3 R_{\rm e,galaxy}$). 
We exclude the inner $1\farcm7$ due to confusion between fine structures and under-subtracted regions. 
The area of the fine structures is $16\%$ of the total area in the range $1\farcm7<R<2.5R_{\rm e,GCS}$. 
We found 32 GC candidates superimposed on these fine structures, which is about $35\%$ of all GCs beyond $1\farcm7$, after background subtraction. 
These numbers are increased when we include less prominent fine structures.
Interestingly, five GC candidates are found on the bright substructure (1) to the east. 

To determine the significance of this positional correlation, we performed a Monte Carlo experiment that produces random GC positions from a smooth GC profile. There are 94 GC candidates with $1\farcm7 < R < 2.5 R_{\rm e,GCS}$. We created random samples of 94 GCs where the position angle, $\theta$, is uniformly random, and $R$ is randomly sampled from the best-fit Sersi\'c profile within the radius limits. 
We did this 100,000 times, and found that only $0.11\%$ of the samples showed an equal or greater number of GCs superimposed on the fine structures than the real data did. 
We also did a same test for the bright substructure (1) at the east, and only $0.09\%$ of results showed an equal or greater number of GCs in the bright substructure than did the real data.
These results suggest that the GCs on the fine structures are physically associated with the fine structures with $\sim99.9\%$ confidence.

\subsection{Colors of GCs in substructures}

\begin{figure}
\epsscale{1.15}
\plotone{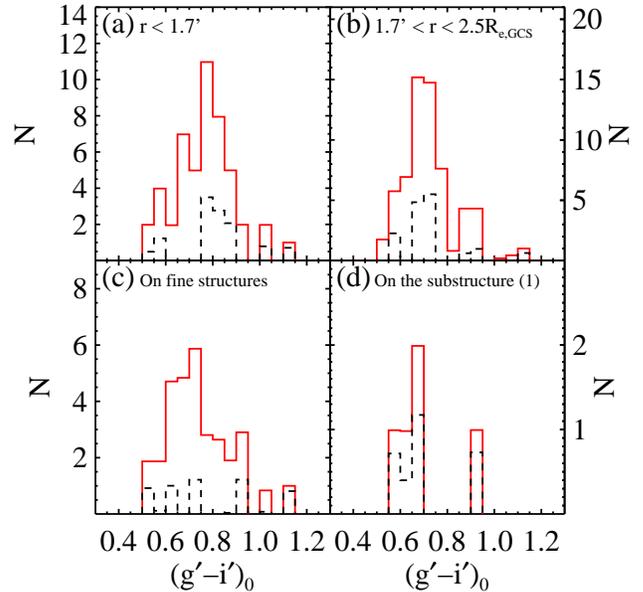}
\caption{Color distributions of GC candidates in different sub-regions. Panel (a), (b), (c), and (d) show GCs within $1\arcmin.7$, $1\farcm7<R<2.5R_{\rm e,GCS}$, fine structures, and the eastern bright substructure, respectively. 
Solid histograms (red) are background subtracted color distributions. 
Dashed histograms (black) are subtracted the double Gaussian function fit to the color distribution of all GCs within 2.5 Re,GC from the color-histograms of GCs in the sub-regions. 
\label{cgr}}
\end{figure}

Figure~\ref{cgr} displays a color histogram of GC candidates in different sub-regions of NGC~474. 
The colors of GCs in the central region (Figure \ref{cgr}a) have a peak at $(g'-i')_0\sim0.85$~mag, which means that GCs in the central region have mostly intermediate-color. 
To better reveal this intermediate-color population, we subtracted the double Gaussian function fit to the color distribution of all GCs within 2.5 $R_{\rm e, GCS}$ from the color-histograms of GCs in the sub-regions. There is an excess at $(g'-i')_0\sim0.85$~mag with GCs in the central region, although GCs in the other sub-regions show excesses at the blue color ($(g'-i')_0\sim0.7$) or no excess.
GCs in the outer region (Figure~\ref{cgr}b) and on fine structures (Figure~\ref{cgr}c) have similar colors because a large portion of the area is the same. In both cases, blue GCs are dominant. 
The colors of GCs in the fine structures are slightly bluer than the mean of the blue peak for the total population (shown by the Gaussian curves).
GCs in the bright substructure (1) are mostly blue, $(g'-i')_0\sim0.7$~mag (Figure~\ref{cgr}d).

\section{Discussion \& Conclusion}

In this study, we found that there is a significant correlation between the GC spatial distribution and the locations of low surface brightness fine structures. Moreover, we find that the GCs superimposed on the fine structures are mostly blue. At least $\sim35\%$ of outer GCs are visibly associated with recent merging.
The bright substructure (1) at the East side of NGC 474 is a good candidate to be the remnant of an accreted dwarf galaxy.  
We roughly estimate the magnitude and color of substructure (1) to be $g'\sim15.5$~mag and $(g-i)_0\sim 0.6$, respectively. 
The color of this substructure is similar to the peak color of GCs in it, suggesting that this substructure (1) and the GCs on it have similar stellar populations. 
The absolute magnitude of this substructure is about $M_{g'}\sim-17.0$, which is on the luminous side for a dwarf galaxy, but still sub-$L^\star$. 
We obtain a GC specific frequency of $S_N \sim 1$ using the number of GC candidates on this feature. This $S_N$ value is a typical value for galaxies with a similar luminosity \citep{Pen08}, so it may support the idea that the GC candidates that lie in projection on this substructure are actually physically associated with the substructure.

We have also found intermediate-color GCs in the central region. 
Intermediate-color GCs in merger remnant galaxies are mostly explained as intermediate-age GCs (e.g. \citealp{Tra14}). 
The ATLAS$^{\rm 3D}$ survey revealed that the mean age of the stellar population in the central region of NGC~474 is $7.65\pm1.39$ Gyr \citep{McD15}, which is younger than the massive cluster early-type galaxies typically studied. The intermediate-color GCs could therefore also have younger ages.
We suggest that the intermediate-color GCs are GCs formed $in-situ$ when star formation occurred at least 7--8 Gyr ago. 

One interesting result in this study is lack of red GCs. 
We found that the fraction of red GC candidates within $2.5R_{\rm e,GCS}$ is about $20\%$. 
\citet{Pen06}, however, showed that the red GC fraction for a galaxy with similar luminosity to NGC~474 should be about $50\%$. 
\citet{Pen06} only studied the central regions of galaxies, so the relatively small fraction of red GCs may be due to large spatial coverage of this study. 
However, \citet{Sik06}, using HST imaging, showed that the fraction of red GCs is small even in the central region of NGC 474, so the lack of red GCs in NGC 474 seems to be a genuine property of the system.

Gas-rich mergers usually increase the number of metal-rich GCs, but if the merger occurred a few Gyr ago, GCs formed in merging are still intermediate in color, and not yet very red. These intermediate-color GCs will become typical red GCs after several Gyr. 
If we assume that these intermediate-color GCs are 7--8 Gyr old with $(g'-i')_0\sim0.85$, then the colors of these GCs will be $(g'-i')_0\sim0.94$ when they are 13 Gyr old with $Z\sim0.003$ based on theoretical simple stellar population model \citep{Mar05,Bre12}.
This red color is consistent with the peak color of red GC populations. 
When these intermediate-color GCs are 13 Gyr old, the fraction of red GCs will be about $40\%$, a fraction consistent with the results in \citep{Pen06}. 

In this study, we have caught both traditional GC sub-populations in mid-evolution. A population of likely intermediate-age GCs are the metal-rich GCs that will passively redden to have colors similar to those seen in older galaxies. The metal-poor population is seen to be physically associated with cold streams and fine structures in the halo of NGC~474. 
Future studies using the MATLAS survey data will allow us to probe this association across a wide range of galaxy mass and environment.
The properties of GCs associated with visible substructures will be useful for determining the origin of streams and shells in massive galaxies.

\acknowledgments

We thank the anonymous referee for helpful comments that improved the original manuscript. SL and EWP acknowledge support from the National Natural Science Foundation of China through Grant No.\ 11573002, and from the Strategic Priority Research Program, ``The Emergence of Cosmological Structures,'' of the Chinese Academy of Sciences, Grant No. XDB09000105. Based on observations obtained with MegaPrime/MegaCam, a joint project of CFHT and CEA/IRFU, at the Canada-France-Hawaii Telescope (CFHT) which is operated by the National Research Council (NRC) of Canada, the Institut National des Science de l'Univers of the Centre National de la Recherche Scientifique (CNRS) of France, and the University of Hawaii. This work is based in part on data products produced at Terapix available at the Canadian Astronomy Data Centre as part of the Canada-France-Hawaii Telescope Legacy Survey, a collaborative project of NRC and CNRS.

{\it Facility:} \facility{CFHT}

\clearpage
\begin{table*}
\small
\begin{center}
\caption{A catalog of the globular clusters in NGC 474 \label{tbl1}}
\begin{tabular}{lllcccc}
\tableline\tableline ID & R.A. & Dec. & $g_0$ & $(u-g)_0$ &
 $(g-r)_0$ & $(g-i)_0$  \\
   & (J2000) & (J2000) & [mag] &  &  & \\
\tableline
121 &  20.05616177 &   3.48833758 & 19.00 $\pm$  0.00 &  2.37 $\pm$  0.00 &  0.65 $\pm$  0.00 &  0.89 $\pm$  0.00 \\
 33 &  19.99084420 &   3.43659182 & 21.26 $\pm$  0.00 &  2.18 $\pm$  0.01 &  0.59 $\pm$  0.00 &  0.92 $\pm$  0.00 \\
 64 &  20.01865022 &   3.33998520 & 21.53 $\pm$  0.00 &  1.36 $\pm$  0.01 &  0.38 $\pm$  0.01 &  0.56 $\pm$  0.01 \\
 36 &  19.99918751 &   3.39075644 & 21.71 $\pm$  0.00 &  1.76 $\pm$  0.02 &  0.48 $\pm$  0.01 &  0.76 $\pm$  0.01 \\
  6 &  19.94028364 &   3.43017381 & 21.96 $\pm$  0.01 &  1.69 $\pm$  0.02 &  0.47 $\pm$  0.01 &  0.70 $\pm$  0.01 \\

\tableline
\tablenotetext{1}{(This table is available in its entirety in a machine-readable form in the online journal. A portion is shown here for guidance regarding its form and content.)}
\end{tabular}
\end{center}
\end{table*}

\end{document}